\documentclass[aip,reprint,pop]{revtex4-1}

\usepackage{graphicx}
\usepackage{subfigure}
\usepackage{epstopdf}
\usepackage{amsmath}
\usepackage{nicefrac}
\usepackage{dcolumn}
\usepackage{bm}
\usepackage{mathrsfs}
\usepackage[colorlinks=true, linkcolor=blue]{hyperref}

\begin{document}

\title{Kinetic Theory of Transport Driven Current in Centrally
fuelled Plasmas}
\author{J.M. Rax}
\affiliation{LOA-ENSTA, Universit\'{e} de Paris XI-Ecole Polytechnique,
91128 Palaiseau, France}
\author{J. Robiche}
\affiliation{LOA-ENSTA, Universit\'{e} de Paris XI-Ecole Polytechnique,
91128 Palaiseau, France}
\author{R. Gueroult}
\affiliation{LAPLACE, Universit\'{e} de Toulouse, CNRS, 31062 Toulouse, France}
\author{C. Ehrlacher}
\affiliation{IRFM-CEA, Cadarache, 13108 Saint Paul lez Durances Cedex, France}

\begin{abstract}
When a steady-state cylindrical plasma discharge is centrally fuelled, the
collisionless radial electron flux is canonically coupled to an axial
current. The identification and analysis of this transport driven current,
previously reported in collisionless simulations [W. J. Nunan and J. M.
Dawson, Phys. Rev. Lett. \textbf{73}, 1628 (1994)], is addressed
analytically and extended to the collisional regime by means of first-principles
kinetic models. Collisionless radial transport is described with the
standard quasilinear model and collisional velocity anisotropy relaxation
with the Landau kinetic equation. When trapped particles corrections are
taken into account, the solution of this kinetic model provides the
analytical expression for the transport driven current in a centrally fuelled
steady-state tokamak as a function of the thermonuclear power and discharge
parameters. For ITER type discharges, with central fuelling, a current of about one mega-ampere is predicted by this first-principles analytical
kinetic model.
\end{abstract}

\date{\today}
\maketitle

\section{Introduction}
\label{Sec:Sec1}

Steady-state tokamak operation displays many practical and economical
advantages~\cite{Wesson2011,Rax2011} and\ a detailed understanding of the various current
generation mechanisms~\cite{Fisch2014} is needed for the successful realization of a
steady-state reactor. The international ITER project in Cadarache is aimed at exploring such
steady-state\ operation in the thermonuclear regime. Together with non-inductive current generation~\cite{Fisch1978,Fisch1987}, the bootstrap effect~\cite{Bickerton1971,Kadomtsev1971} is expected
to provide a fraction of its $15$ mega-amperes current. 

Besides the bootstrap
current, another self-generated current, without seed, was observed and
reported~\cite{Nunan1994,Ma1994} in numerical simulations when a magnetized cylindrical
plasma discharge is centrally fuelled. In contributing to the toroidal current which provides the rotational transform required to cancel the vertical drift, this effect could affect plasma operation. Understanding and quantifying this effect is therefore particularly important for steady-state operation in ITER. Yet, this so-called transport driven
current has received far less attention than the bootstrap current. 

In this paper, we set up, solve and analyze an analytical model based on the
quasilinear theory~\cite{Galeev1968,Kaufman1972} to describe turbulence-particle interactions
and on the Landau collisional relaxation~\cite{Robiche2004} to describe
particle-particle interactions. This first-principles model makes it possible to
evaluate the expected fraction of spontaneous transport driven current in a
centrally fuelled ITER type discharge when trapped particles corrections are
taken into account. As it will be shown, it turns out that a non-negligible fraction of the
confining current, about a mega-ampere, is expected to result from this
overlooked effect.

Non-inductive current generation at a level of tens of mega-amperes is
needed for a steady-state reactor~\cite{Fisch2014,Fisch1978,Fisch1987} and additional spontaneous
toroidal currents, like the bootstrap~\cite{Bickerton1971,Kadomtsev1971} and the effect analyzed in
this paper~\cite{Nunan1994,Ma1994}, will improve the global power balance of a burning
thermonuclear plasma.

Up to now, large tokamak discharges have been operated via edge gas puffing
or pellet fuelling~\cite{Wesson2011,Rax2011}. In comparison with edge or mid-radius fuelling, central fuelling offers conceptual advantages to
peak the pressure profile and to wash out ashes and impurities which naturally accumulate near the magnetic axis. Although, it is not clear
how to perform efficiently central fuelling in reactor size plasma, it is
important to assess the physical consequences of central fuelling on a
steady-state reactor.

The transport driven current in cylindrical or toroidal configurations is
associated with the radial flux of charges in a centrally fuelled discharges.
The favorable and encouraging result presented here provides a supplementary drive to
develop fuelling systems able to deposit fuel very near the core of the
discharge.

For a burning plasma where deuterium (D) and tritium (T) are continuously
deposited near the magnetic axis, and helium ashes (He) continuously removed
in the scrape of Layer (SOL) through the divertor, we estimate that the driven
current is in the mega-ampere range for typical reactor size and performances.
This new result is based on the following assumptions: (\textit{i}) a
steady-state burning D-T plasma where (\textit{ii}) electron transport is
due to turbulence in the quasilinear regime, (\textit{iii}) current
relaxation is due to electron/ion collisions, (\textit{iv}) collisional
transport is negligible in comparison with turbulent transport and (\textit{v}) the
turbulent modes are associated with rational magnetic surfaces. These five
assumptions are canonical for standard tokamak models and provide a clear
and realistic framework for this analytical study.

\begin{figure}
\begin{center}
\includegraphics[width=7cm]{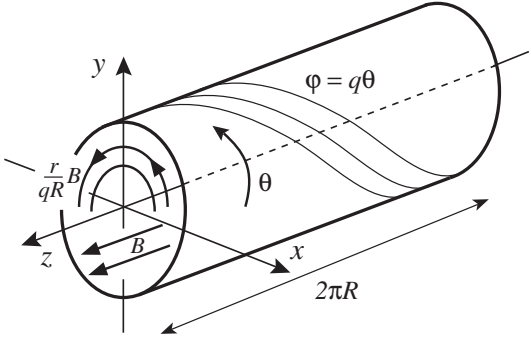}
\caption{A straight tokamak configuration with minor radius $a$, major radius $R$ and safety factor $q$.}
\label{Fig:Fig1}
\end{center}
\end{figure}

Consider a straight tokamak reactor model (a cylindrical screw pinch
illustrated in Fig.~\ref{Fig:Fig1}) with a global fusion power $W_{f}$ ($\sim $ GW) . To
discriminate the central fuelling effect from the bootstrap effect, or other neoclassical effects, we
restrict the model to a straight tokamak with minor radius $a$, major radius 
$R$, safety factor $q$ and typical magnetic field $B$. The impact of trapped
particles on transport driven current in tokamak is evaluated in the last
section. In order to operate in steady-state, a radial flux of matter is
needed from the center, where fuelling and combustion take place, toward the
edge, where ashes and heat removal are operated in the SOL.

If neutral fuel is deposited and burned near the magnetic axis, the radial
electrons flux $\Gamma _{r}$ at radius $r$ is given by
\begin{equation}
\Gamma _{r}=\frac{W_{f}}{4\pi ^{2}rRQ_{DT}},
\end{equation}
where $Q_{DT}$ is the energy yield per D/T fusion reaction ($17.6$ MeV). In
this model the rate of particles central-injection/edge-extraction, $%
W_{f}/Q_{DT}$, corresponds to a radial ambipolar flux. However, because of their
charge to mass ratio, ions are only involved through pitch-angle scattering
current destruction and not through current generation. This net outward
flux $\Gamma _{r}$, independent of the recycling processes~\cite{Wesson2011,Rax2011}, is a
consequence of the steady-state and central fuelling requirements and its
precise nature, be it convective~\cite{Ware1970}, diffusive or even non local~\cite{Rax1989},
remains an open question. A mean\ radial electron velocity $V$ is associated
with this flux:
\begin{equation}
V=\frac{\Gamma _{r}}{n_{e}}\approx \frac{W_{f}}{4\pi
^{2}aRQ_{DT}\left\langle n_{e}\right\rangle },
\end{equation}
where $n_{e}$ is the electron density and $\left\langle n_{e}\right\rangle $
is the electron density averaged over the whole discharge volume. 

Electrons transport from the core towards the SOL is expected to take place in the
turbulent regime associated with a spectrum of modes. Let us consider one
turbulent mode such that its local structure is periodic along the magnetic
field line, with wavelength $2\pi /k_{\Vert }$, and periodic across the
field lines, in the poloidal direction, with wavelength $2\pi /k_{\perp }$.
Under the random phase approximation (RPA)~\cite{Rax2014}, such a $\left( k_{\Vert },k_{\perp }\right)$ electrostatic, or
electromagnetic, wave interacting with an electron transfers linear momentum $m_{e}\delta v_{\Vert }$ along the magnetic field. It also induces a displacement $\delta r$ of the
guiding center radial position, which is related to this momentum transfer by
\begin{equation}
\Omega k_{\Vert }\delta r=\delta v_{\Vert }k_{\perp },
\end{equation}
where $\Omega =eB/m_{e}$ is the cyclotron frequency, $e$ and $m_{e}$ the
electron charge and mass. This effect is due to canonical momentum conservation and is put at work in alpha particles free-energy extraction~\cite{Fisch1992,Fisch1993,Fisch1994,Heikkinen1995,Herrmann1997,Ochs2015,Cook2017} and in
angular momentum injection in advanced tokamaks~\cite{Rax2017}.

The linear momentum increment $m\delta v_{\Vert }$ given by the wave (which
becomes a toroidal angular momentum in a tokamak configuration) is then
dissipated through collisions, mainly through ion pitch-angle scattering,
at a rate $\nu $, the pitch-angle scattering collision frequency~\cite{Wesson2011,Rax2011}.
The small transient toroidal current $\delta I$ is thus given by the
expression: 
\begin{align}
\delta I\left( \delta r,t\right) = & \frac{e}{2\pi R}\delta v_{\Vert }\exp
\left( -\nu t\right) \nonumber\\ = &~\Omega \frac{k_{\Vert }}{k_{\perp }}\frac{e}{2\pi R}%
\delta r\exp \left( -\nu t\right) \text{.}
\end{align}
We do not need the full picture of the transport process, convective,
diffusive, local or non local, because we know that the sum of all the
stochastic radial steps $\delta r$ is ultimately given by: $\Sigma \delta
r=a$, when the center to edge transit is achieved, even with strong
recycling. We also know that the average rate of radial transport $%
\left\langle \delta r/\delta t\right\rangle $ must be equal to $V$ in steady-state. The steady-state current $\left\langle \delta I\right\rangle $ is
given by the sum of the time-averaged incremental currents created by each
electron: $\left\langle \delta I\right\rangle =\int \delta Idt/\int dt$
where $\int dt=\delta r/V$ is the time needed for one radial step $\delta r$. Therefore, for each electron 
\begin{equation}
\left\langle \delta I\right\rangle =\frac{\int_{0}^{\infty }\delta I\left(
t\right) dt}{\delta r/V}=\frac{\Omega }{\nu }\frac{k_{\Vert }}{k_{\perp }}%
\frac{eV}{2\pi R}\text{.}
\end{equation}
This one electron and one wave result must be multiplied by the total number
of electrons $2\pi ^{2}a^{2}R\left\langle n_{e}\right\rangle $ and averaged
over the full turbulence spectrum. Thus, for a reactor with power $W_{f}$ and
mean spectral characteristics $\left\langle k_{\Vert }/k_{\perp
}\right\rangle $ we get the final current estimate 
\begin{equation}
I\approx \frac{\Omega }{4\pi \nu }\frac{a}{R}\left\langle \frac{k_{\Vert }}{%
k_{\perp }}\right\rangle \frac{eW_{f}}{Q_{DT}}\text{.}
\end{equation}
For a tokamak discharge the ratio of the local wave numbers $k_{\Vert
}/k_{\perp }$ can be expressed as a function of the modes numbers $m$ and $n$
: $m$ is the poloidal mode number ($\theta $ is the poloidal angle) and $n$
is the toroidal mode number ($\varphi $ is the toroidal angle) such that $%
k_{\Vert }/k_{\perp }=$ $nr/mR\sim $ $na/mR$.

The hypothesis of resonant modes localized near closed field lines, as a
result of magnetic shear for drift types modes~\cite{White2014}, implies that $m+nq\sim
0 $ where $q$ is the safety factor associated with a closed helical field
line, $\varphi =q\theta $. With this rough estimate, $\left\langle
k_{\Vert }/k_{\perp }\right\rangle \sim $ $a/qR$, the 
typical transport driven current for a centrally fuelled
thermonuclear reactor with power $W_{f}$ $\ $is
\begin{equation}
\frac{I}{W_{f}}\approx \frac{\Omega }{\nu }\frac{1}{4\pi q}\frac{a^{2}}{R^{2}%
}\frac{e}{Q_{DT}}\sim 10^{-9}\frac{\Omega }{\nu }\left[ \frac{\text{A}}{%
\text{W}}\right],
\end{equation}
where $\left( Q_{DT}/e\right) =17,6\times
10^{6}$ [Joule/Coulomb] has been assumed for the D/T reaction and the rough estimates $2\pi q\sim 10$ and $\left( R/a\right) ^{2}\sim
10 $ have been used for an ITER type discharge. Despite the $10^{-9}$ factor, the transport driven current $I$ is not
negligible in a burning centrally fuelled discharge since (\textit{i}) the
fusion power $W_{f}\sim 10^{9}$ Watt and (\textit{ii}) typically $\Omega
\geq 100$ GHz and $\nu \leq 1$ MHz such that $\Omega /\nu \sim 10^{6}$. This
simple estimate leads to a current of about a 
mega-ampere, which is comparable to the expected contribution of the bootstrap current. It is worth noting here that this effect is not associated with the asymmetry of
the poloidal spectrum $\left\langle m\right\rangle $, or the toroidal
spectrum $\left\langle n\right\rangle $, but with the finite value of the
mean value of the ratio $\left\langle n/m\right\rangle \sim 1/q$. 

The encouraging prediction of this heuristic model will be validated by laying out and solving a full quasilinear-kinetic model in
the next sections.

The interaction between an electron and a spectrum of electrostatic modes
with the space $\left( \theta ,\varphi \right) $ and time $\left( t\right) $
structure $\exp j\left( m\theta +n\varphi -\omega t\right) $ is considered in
Sec.~\ref{Sec:Sec2} under the RPA approximation in order to set up the classical
quasilinear picture. By considering a straight tokamak with safety factor $q$, the
turbulent spectrum is expanded on a Bessel cylindrical basis. The
analysis is developed with the angle-action variables to separate
slow and fast motions and to average over the fast phase according to the
RPA prescription.

The result of this Hamiltonian quasilinear analysis is then used in Sec.~\ref{Sec:Sec3} to construct a collisional relaxation model describing the main current
dissipation mechanism: pitch-angle scattering. The collisional Landau
kinetic equation is solved with a Legendre polynomial expansion and the
current associated with turbulent RPA transport is derived. 

As a conclusion,
Sec.~\ref{Sec:Sec4} presents a discussion on the validity and limits of this analytical
fully-kinetic model and explores the implications of this new result for
centrally fuelled ITER type discharges.

\section{Quasilinear analysis of turbulence-electron interaction}
\label{Sec:Sec2}

Let us consider a straight tokamak configuration as illustrated in Fig.~\ref{Fig:Fig1}. The magnetic field $\mathbf{B}$ of this screw pinch can be decomposed as the sum
of a two components: a toroidal component along the $z$ magnetic axis plus a
poloidal component, increasing linearly from the center toward the edge and typically
smaller by a factor $a/qR$,
\begin{equation}
\frac{e}{m_{e}}\mathbf{B}=\Omega \mathbf{e}_{z}-\frac{\Omega }{qR}y\mathbf{e}%
_{x}+\frac{\Omega }{qR}x\mathbf{e}_{y}\text{,}  \label{mag}
\end{equation}
where $\left[ x\mathbf{,}y\mathbf{,}z\right] $ is a set of Cartesian
coordinates (see Fig.~\ref{Fig:Fig1}) and $\left[ \mathbf{e}_{x},\mathbf{e}_{y},\mathbf{e}%
_{z}\right] $ a Cartesian basis.

The orbit of an electron confined by this magnetic configuration is the
combination of a fast cyclotron rotation around the field line plus a fast
translation along the field lines. If the toroidal curvatures effects are
taken into account the slow vertical drift across the field lines is
cancelled by the poloidal rotation. This magnetic configuration with helical
magnetic fields Eq.~(\ref{mag}) is described by the vector potential 
\begin{equation}
\frac{e}{m_{e}}\mathbf{A}=-\Omega y\mathbf{e}_{x}-\frac{\Omega }{qR}\frac{%
x^{2}+y^{2}}{2}\mathbf{e}_{z}\text{.}
\end{equation}
The Hamiltonian $H_{0}$ of an electron with canonical momentum $\mathbf{p}$
is thus given by~\cite{Kaufman1972,White2014}
\begin{equation}
H_{0}=\frac{\left( \mathbf{p-A}\right)^{2}}{2}=\frac{1}{2}\left( \mathbf{p}+\Omega y\mathbf{e}%
_{x}+\frac{\Omega }{qR}\frac{x^{2}+y^{2}}{2}\mathbf{e}_{z}\right) ^{2}%
\text{,}  \label{ham}
\end{equation}
where we have normalized the unit of charge and electron mass $e=m_{e}=1$.

\begin{figure*}
\begin{center}
\includegraphics[width=10cm]{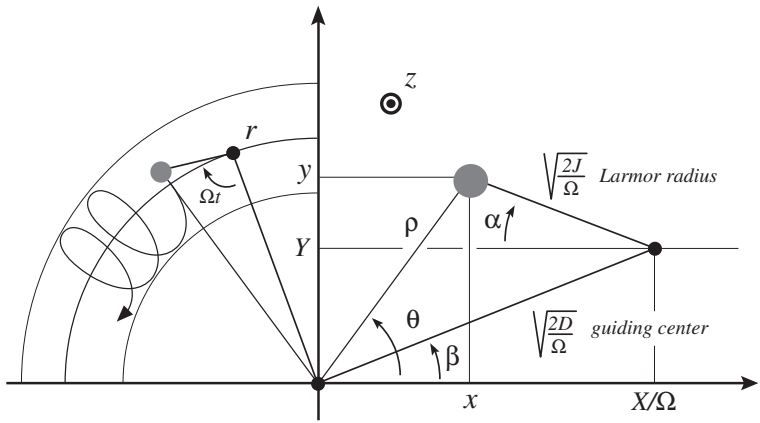}
\caption{Guiding center $\sqrt{2D/\Omega}$ and Larmor radius $\sqrt{2J/\Omega}$ action variables in a poloidal cross section $(x, y)$. }
\label{Fig:Fig2}
\end{center}
\end{figure*}

In order to analyze quasilinear transport, we first perform a canonical
transform from the $\left( p_{x},p_{y}\right)$ and $\left( x,y\right)$ old
poloidal variables to the new actions $\left( J,X\right) $ and angles $%
\left( \alpha ,Y\right) $ with the generating function~\cite{Rax2011} of the first
type 
\begin{equation}
F_{1}\left( x,y,\alpha ,Y\right) =\Omega \left( y-Y\right) ^{2}\cot
\left( \alpha \right) /2-\Omega xY
\end{equation}
illustrated in Fig.~\ref{Fig:Fig2}. The final
result is given by the classical set of relations: $x=X/\Omega -\sqrt{%
2J/\Omega }\cos \alpha $ and $y=Y+\sqrt{2J/\Omega }\sin \alpha $ where the
geometrical meaning of the guiding center variables $X/\Omega $ and $Y$ are
displayed in Fig.~\ref{Fig:Fig2}.

Then, rather than this Cartesian guiding center variables $X/\Omega $ and $Y$, we will use the polar variables $D$ and $\beta $ obtained through a second
canonical transform generated by the generating function of the first type 
\begin{equation}
G_{1}\left( Y,\beta \right) =\Omega Y^{2}\cot \left( \beta \right) /2.
\end{equation}
The final set of guiding center actions $\left(J,D\right) $ and angles $\left(
\alpha ,\beta \right) $ variables, illustrated in Fig.~\ref{Fig:Fig2}, can be interpreted
as the guiding center polar coordinates $(r=\sqrt{2D/\Omega},\beta) $ and the Larmor radius and cyclotron angle $(\sqrt{2J/\Omega},\alpha) $. They are related to the electron poloidal position $%
\left( x,y\right) $ through 
\begin{align}
x = &\sqrt{\frac{2D}{\Omega }}\cos \beta -\sqrt{\frac{2J}{\Omega }}\cos \alpha,\nonumber\\ y = &\sqrt{\frac{2D}{\Omega }}\sin \beta +\sqrt{\frac{2J}{\Omega }}%
\sin \alpha.
\end{align}
The radial electron coordinate $\rho =\sqrt{x^{2}+y^{2}\text{ }}$can then be
written as a function of the angles-actions variable
\begin{equation}
\frac{\Omega }{2}\rho ^{2}=D+J-2\sqrt{JD}\cos \left( \alpha +\beta \right).
\end{equation}

Introducing the canonical momentum along the magnetic field
line $P=\mathbf{p}\cdot \mathbf{e}_{z}$, conjugate to the $z$ variable, we
can express the Hamiltonian $H_{0}$ Eq.~(\ref{ham})\thinspace as 
\begin{align}
H_{0} = &~\Omega J+\frac{P^{2}}{2}+\frac{P}{qR}\left[ D+J-2\sqrt{JD}\cos
\left( \alpha +\beta \right) \right] \nonumber\\
 & +\frac{1}{2q^{2}R^{2}}\left[ D+J-2\sqrt{JD}\cos \left( \alpha +\beta
\right) \right] ^{2}.
\end{align}

The guiding center poloidal rotation resulting from the helical structure of
the field lines is much slower than the cyclotron rotation: $d\alpha
/dt=\partial H_{0}/\partial J$ $\approx \Omega $ $\gg d\beta /dt=\partial
H_{0}/\partial D$ $\approx P/qR$. This strong ordering allows to safely
average the oscillating terms over $\alpha $ as no resonance between $\alpha 
$ and $\beta $ can take place. The adiabatic Hamiltonian $H_{0}$ describing
the guiding center orbits in this adiabatic screw-pinch/straight-tokamak
configuration is
\begin{equation}
H_{0}=\Omega J+\frac{P^{2}}{2}+P\frac{J+D}{qR},  \label{ham2}
\end{equation}
where we have neglected the last term on the right hand side since $a/qR<1$.

The physics behind this Hamiltonian is rather simple: (\textit{i}) $\Omega
J $ is the cyclotron rotation energy around the field lines, (\textit{ii}) $%
P^{2}/2$ the translation energy along the $z$ direction and (\textit{iii}) $%
P\left( J+D\right) /qR$ the kinetic energy associated with the poloidal
rotation due to the helicity of the field line. Without collisions or
turbulence, $H_{0}$ given by Eq.~(\ref{ham2}) describes perfect adiabatic
confinement such that $dD/dt=dJ/dt=dP/dt=0$.

Within the framework of the turbulent transport driven current problem we
are interested by the coupled dynamics of the guiding center radial position 
$\sqrt{2D/\Omega }$ and the momentum $P$. This coupling is induced by a
spectrum of turbulent modes. We will only consider here electrostatic
modes and point out that electromagnetic modes described by a perturbating vector potential included in Eq.~(\ref{ham}) would yield the same final result. 

Consider an electrostatic turbulent spectrum described by the scalar potential 
\begin{widetext}
\begin{equation}
\Phi \left( r,\theta ,z,t\right) =\sum_{m=-\infty }^{m=+\infty
}\sum_{n=-\infty }^{n=+\infty }\Phi _{mn}\left( r\right) \exp \left(
jm\theta \right) \exp \left( jn\frac{z}{R}\right) \exp \left( -j\omega _{mn}t\right) 
\end{equation}
\end{widetext}
where $\Phi _{mn}\left(
r\right) $ is the radial eigenmode associated with the poloidal and toroidal mode $\exp (jm\theta) \exp (jn\varphi)$, with $\theta $ and $\varphi $ ($R\varphi =z$) the poloidal
and toroidal angle of the straight tokamak (see Fig.~\ref{Fig:Fig1}), for a given frequency $\omega _{mn}\ll \Omega $.
The structure of the radial eigenmode $\Phi _{mn}\left( r\right) $ is very
important to set up the various model of tokamak instabilities and
turbulence, but it is not needed to derive kinetic theory of transport driven
current. We simply assume that it can be decomposed on a
natural cylindrical basis of ordinary Bessel functions of order $m$, $%
J_{m}\left( kr\right) $, and the $k$ Fourier variable can be discrete
(Fourier series) if we impose a boundary condition at $r=a$, or continuous otherwise
(Fourier integral). The precise nature of this radial expansion does not
change the final results of this analytical model of current generation. We thus consider a classical Fourier-Bessel expansion providing a simple
identification of resonant transport~\cite{Watson1980}: 
\begin{align}
\Phi _{mn}\left( r\right) = & \int_{0}^{+\infty }kdk\phi _{mn}\left( k\right)
J_{m}\left( kr\right); \nonumber\\ \phi _{mn}\left( k\right)
= & \int_{0}^{+\infty }rdr\Phi _{mn}\left( r\right) J_{m}\left( kr\right) \text{%
.}
\end{align}
The random phase approximation (RPA) assumes that the effect of each mode
can be analyzed separately within the Hamiltonian framework and that the
full quasilinear effect is just the sum of these single mode perturbations
on the actions averaged over the angles (RPA)~\cite{Wesson2011,Rax2014,White2014}. Within this
canonical framework let us consider the Hamiltonian $H$ describing the
interaction between one electron and one $\left( k,m,n\right) $ mode: 
\begin{multline}
H = ~\Omega J+\frac{P^{2}}{2}+P\frac{J+D}{qR} 
\\ +\phi _{mn}\left( k\right)
J_{m}\left( kr\right) \exp(j[m\theta+n\varphi-\omega _{mn}t]).
\end{multline}
This RPA-quasilinear analysis can be further simplified with the help of the
Gegenbauer's addition theorem~\cite{Watson1980} 
\begin{multline}
J_{m}\left( kr\right) \exp (jm\theta) =  \sum_{l=-\infty }^{l=+\infty
}J_{l+m}\left( k\sqrt{\frac{2D}{\Omega }}\right) J_{l}\left( k\sqrt{\frac{2J%
}{\Omega }}\right) \\ \times\exp (jl\alpha) \exp \left[j\left( l+m\right) \beta\right].
\end{multline}
This final derivation reduces the analysis of the $\left( k,m,n\right) $
coupling term to a sum over the integer $l$ associated with the order of the
cyclotron resonance. For typical electrostatic turbulence $\omega _{mn}\ll
\Omega $, so fundamental $l=1$, anomalous $l=-1$ and harmonic $\left|
l\right| >1$ cyclotron resonant interactions do not take place. We can then
neglect all the components $l\neq 0$ and restrict the model to the $l=0$ $\ $%
component. The low frequency Hamiltonian describing the coupling between the 
$\left( k,l=0,m,n\right) $ mode and one electron is thus given by
\begin{multline}
H =  H_{0}+\phi _{mn}\left( k\right) J_{m}\left( k\sqrt{\frac{2D}{\Omega }}%
\right) J_{0}\left( k\sqrt{\frac{2J}{\Omega }}\right) \\   \times\exp (jm\beta) \exp
(jn\varphi) \exp (-j\omega _{mn}t).  \label{ham3}
\end{multline}
In order to write Hamilton's equations and to display the breakdown of
adiabatic confinement leading to the occurrence of radial transport, we
introduce the phase $\Psi _{mn}=m\beta +n\varphi -\omega _{mn}t$ $\ $and its
unperturbed evolution $\Psi _{mn}\approx \varpi_{mn} t$ with $\varpi_{mn} = mP/qR+nP/R-\omega
_{mn}$ to write Hamilton's equations:
\begin{align}
\frac{dH}{dt} = & \frac{\partial H}{\partial t} = -j\omega _{mn}\phi
_{mn}\left( k\right) \mbox{\large $\Xi$}  \exp (j\Psi _{mn}),\\
\frac{dD}{dt} = & -\frac{\partial H}{\partial \beta }=-jm\phi _{mn}\left(
k\right) \mbox{\large $\Xi$}  \exp (j\Psi _{mn}),\\
\frac{dP}{dt} = & -\frac{\partial H}{R\partial \varphi }=-j\frac{n}{R}\phi
_{mn}\left( k\right) \mbox{\large $\Xi$}  \exp (j\Psi _{mn}),
\end{align}
with
\begin{equation}
\mbox{\large $\Xi$} = J_{m}\left( k\sqrt{\frac{2D}{\Omega }}\right)
J_{0}\left( k\sqrt{\frac{2J}{\Omega }}\right).
\end{equation}
Considering the turbulent term $\phi _{mn}$ in Eq.~(\ref{ham3}) as a
perturbation of the adiabatic Hamiltonian $H_{0}$ \ from Eq.~(\ref{ham2}) we
can integrate these equations during a small time $\delta t$, larger than
the period of oscillations of the angles but smaller than the quasilinear
evolution of the distribution function in actions space $F\left(
J,D,P,t\right) $, in order to get the short time evolution of the energy $H$, the guiding center radial position $D$ and the momentum $P$:
\begin{align}
\delta H_{kmn}\left( \delta t\right)  =&-\omega _{mn}\frac{\phi
_{mn}}{\varpi_{mn}}\mbox{\large $\Xi$}\exp (j\varpi_{mn} \delta t),  \label{ql1} \\
\delta D_{kmn}\left( \delta t\right)  =&-m\frac{\phi _{mn}}{\varpi_{mn}}\mbox{\large $\Xi$}\exp (j\varpi_{mn} \delta t),  \label{ql2} \\
\delta P_{kmn}\left( \delta t\right)  =&-\frac{n}{R}\frac{\phi
_{mn}}{\varpi_{mn}}\mbox{\large $\Xi$}\exp (j \varpi_{mn} \delta t).  \label{ql3}
\end{align}

The distribution function in action space at time $t$, $F\left(
J,D,P,t\right) $, is the solution of a diffusion equation, the quasilinear equation. The diffusion coefficients of the quasilinear equation are given by the sum over $k$, $m$ and $n$ in Fourier space of the RPA averages
\begin{subequations}
\begin{align}
& \frac{\left\langle \delta
 {D_{kmn}}^{2}\right\rangle}{2\delta t},\\
& \frac{ \left\langle \delta
{P_{kmn}}^{2}\right\rangle}{2\delta t},
\end{align}
and
\begin{align}
& \frac{\left\langle \delta
{D_{kmn}}\delta P_{kmn}\right\rangle}{\delta t}
\end{align}
\end{subequations}
(we should add a $%
j\varepsilon $ to the $\varpi_{mn}$ in the denominator to account for
causality starting from the past $t=-\infty $). However, there is no need to carry out
this standard derivation of the quasilinear theory to derive the kinetic theory of transport driven current
in a centrally fuelled discharge. We only need Eqs.~(\ref{ql1}, \ref{ql2}, \ref{ql3}) to conclude that the ratio of the change of the
radial position of the guiding center, $\delta D_{kmn}$, to the increment of
momentum along the axial/toroidal direction, $\delta P_{kmn}$, does not 
depend on $k$ and takes the simple value $Rm/n$ for an $\left(
m,n\right) $ mode. Indeed, this relation writes
\begin{equation}
n\delta D_{mn}=Rm\delta P_{mn}.
\label{Eq:Eq30}
\end{equation}
If we introduce the radial guiding center position $r=\sqrt{2D/\Omega }$
(Fig. 2) such that $\Omega \delta r^{2}=2\delta D_{mn}$ and the parallel
velocity $v_{\Vert }$ such that $m_{e}\delta v_{\Vert }=$ $\delta P_{mn}$, Eq.~(\ref{Eq:Eq30}) rewrites
\begin{equation}
n\Omega r\delta r=mR\delta v_{\Vert }, 
\end{equation}
which is
similar to the heuristic result derived in the introduction. This
straightforward and general result is the starting point of the collisional
kinetic analysis of the steady-state.

\section{Kinetic collisional theory of current relaxation}
\label{Sec:Sec3}

Both the heuristic approach presented in the introduction, and the more
rigorous Hamiltonian/RPA theory of section two Eqs.~(\ref{ql1}, \ref{ql2}, 
\ref{ql3}), lead to the following conclusion: if a low-frequency turbulent
mode with poloidal number $m$ and toroidal number $n$ yields a guiding
center radial kick $\delta r$, the elementary step of quasilinear diffusion,
then a velocity kick $\delta v_{\Vert }$ is associated with this incremental
radial transport: 
\begin{equation}
\delta v_{\Vert }=\frac{r}{R}\frac{n}{m}\Omega \delta r\text{.}  \label{step}
\end{equation}
This fundamental property, used in free energy extraction~\cite{Fisch1992,Fisch1993,Fisch1994,Heikkinen1995,Herrmann1997,Ochs2015,Cook2017} and
angular momentum injection for advanced tokamak~\cite{Rax2017}, allows to set up the
following physical picture for turbulent transport in a centrally fuelled
discharge: an electron starts on the magnetic axis a random walk towards the
edge. For every step $\pm \delta r$ it takes along this random walk
under the influence of an $\left( m,n\right) $ mode, it gains or looses an
incremental momentum $\delta v_{\Vert }$.

\begin{figure*}
\begin{center}
\includegraphics[width=12cm]{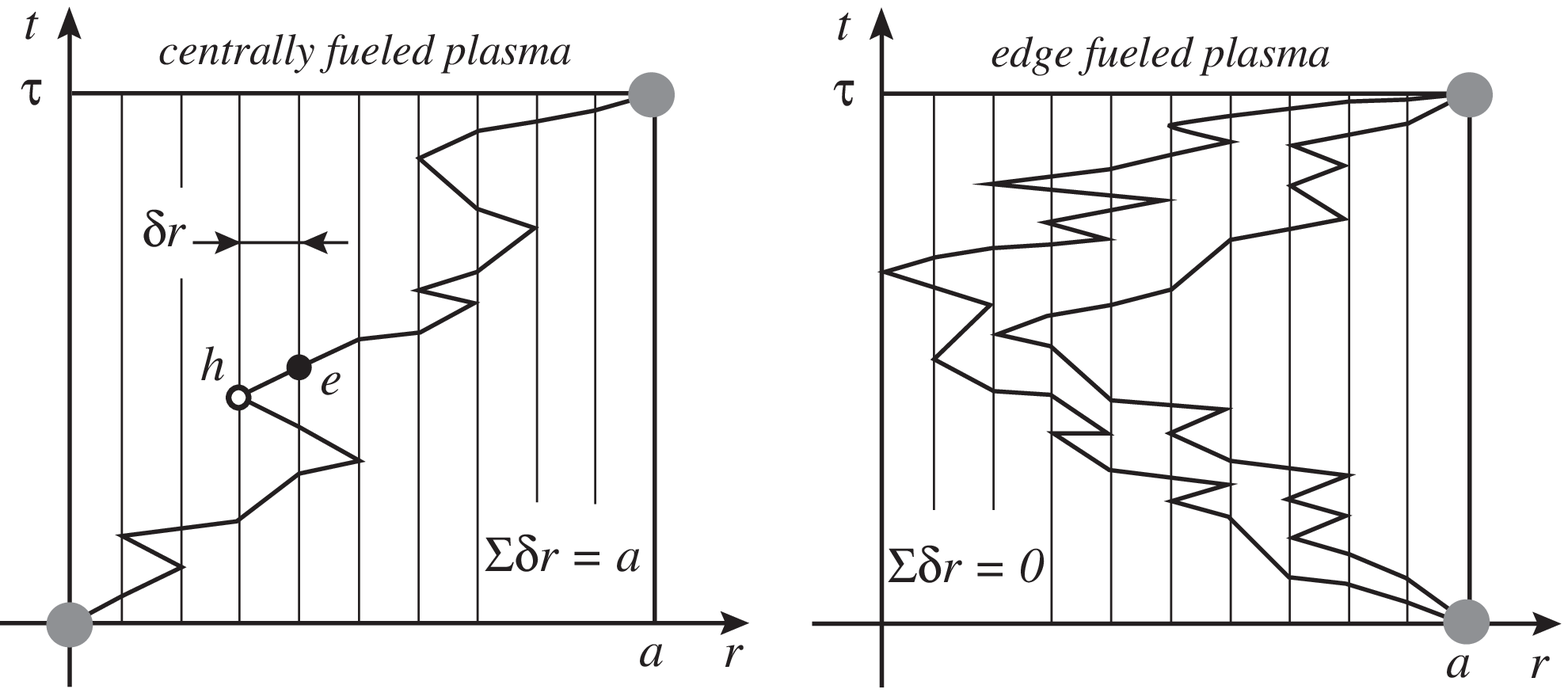}
\caption{Random walk of an electron from a drift surface $r$ to the next drift surface $r + \delta r$ with central and edge fuelling. }
\label{Fig:Fig3}
\end{center}
\end{figure*}

We now recall the concept of electron and hole~\cite{Rax1988}. An electron with velocity $v_{\Vert }$, located on the
drift surface at radius $r$, jumps on a neighboring drift surface at $r+\delta
r$. This is the basic step of the quasilinear random walk. This basic step
creates a hole ($h$) in the distribution function at $\left( r,v_{\Vert
}\right) $ and an additional electron ($e$) at $\left( r+\delta r,v_{\Vert
}+\delta v_{\Vert }\right) $. This electron/hole picture of the quasilinear
random walk is shown in Fig.~\ref{Fig:Fig3} and has been already used to calculate
the non-inductive current efficiency~\cite{Fisch1987,Rax1988}. Fig.~\ref{Fig:Fig3} also illustrates the main difference between an edge fuellled and a centrally fuelled discharge. For edge fuelling, the sum of random kicks or radial transport, $\sum_{\text{random walk}}\delta r$, is equal to zero. In contrast, in a centrally fuelled discharge, $\sum_{\text{random
walk}}\delta r=a$.

Tokamak experimental results show that the electron
population is thermalized and isotropic on drift surfaces, so we
consider that collisional thermalization and isotropization are fast
processes $\left( \sim 10^{-2}s\right) $ compared with radial transport $%
\left( \sim 1s\right) $. This strong ordering between collisionless radial
transport from drift surface to drift surface and collisional relaxation of
the electron and hole excitations justifies the following assumption. The relaxation of the
electron/hole is considered as a kinetic process whose initial condition are given by $%
\left( r,v_{\Vert }\right) \rightarrow \left( r+\delta r,v_{\Vert }+\delta
v_{\Vert }\right) $ at time $t=0$ and with no interference with a further
quasilinear step $\delta r$ during the isotropization process.

Since we are only interested by the current we can restrict the Landau
collisional kinetic equation to pitch-angle scattering on ions. This
restriction is also used in the kinetic theory of the Spitzer conductivity for inductive current generation
and the kinetic theory of the Fisch efficiency~\cite{Fisch1987} for non-inductive current generation.To study the model of Landau collisional relaxation of one hole at $%
\left( r,v_{\Vert }\right) $ and one electron at $\left( r+\delta r,v_{\Vert
}+\delta v_{\Vert }\right) $ we consider a spherical set of coordinates in
velocity space directed by the $z$ axis and introduce  the pitch-angle of
electrons
\begin{equation}-1\leq \mu =v_{\Vert }/v\leq +1
\end{equation}
where $v=\sqrt{v_{\Vert
}^{2}+v_{c}^{2}}$ is the electron velocity and $v_{c}$ the cyclotron
velocity. The distribution function $f_{e/h}\left( r,v,\mu ,t\right) $
describes the electron/hole dynamics near the drift surface $r$ resulting
from a $\left( \delta r,\delta v_{\Vert }\right) $ step at $t=0$.

According to Eq.~(\ref{step}) for a given $\left( m,n\right) $ turbulent
drive, the evolution of this distribution function $f_{e/h}\left( r,v,\mu
,t\right) $ is constrained to take place along quasilinear diffusion paths
such that: $Rm\delta \left( v\mu \right) =\Omega rn\delta r$. Going back to
the actions evolutions given in Eqs.~(\ref{ql1}, \ref{ql2}, \ref{ql3}), the ratio of
the RPA energy kick Eq.~(\ref{ql1}) to the parallel momentum kick Eq.~(\ref
{ql3}) is given by $\delta H/\delta D=\delta v^{2}/2\Omega r\delta r=\omega
_{mn}/m$. In spherical coordinates $v_{\Vert }=v\mu $ \ and the velocity
space modification associated with a radial step $\delta r$ under the
influence of an $\left( m,n\right) $ mode is thus described by the pitch
angle kick
\begin{equation}
\delta \mu _{mn}=\Omega \frac{r}{R}\frac{\delta r}{v}\left( \frac{n}{m}-\mu 
\frac{\omega _{mn}R}{mv}\right) \text{.}  \label{step2}
\end{equation}
As we will ultimately average over a Maxwellian distribution for$\ v$, we
will not consider the energy slowing down and diffusion and we concentrate
on pitch-angle scattering which preserves $v$ owing to the large ion to electron
mass ratio. 
We
introduce the classical collision time~\cite{Wesson2011} $\tau $ defined as
\begin{equation}
\tau \left( v,r\right) =\frac{8\pi {\varepsilon _{0}}^{2}{m_{e}}^{2}v^{3}}{%
e^{4}n_{e}\left( r\right) \log \Lambda }\text{,}
\end{equation}
and the effective ion charge state $Z$. The fast collisional decay of an
electron-hole excitation $v\mu _{0}\rightarrow v\left( \mu _{0}+\delta \mu
_{mn}\right) $ created at time $t=0$ near $r$ is described by the kinetic
equations~\cite{Fisch1987,Rax1988}
\begin{multline}
\left[ \frac{\partial }{\partial t}-\frac{Z+1}{\tau }\frac{\partial }{%
\partial \mu }\left( 1-\mu ^{2}\right) \frac{\partial }{\partial \mu }%
\right] \text{ }f_{h}\left( r,\mu ,t\right) \\ = -\delta \left( \mu -\mu
_{0}\right) \delta \left( t\right),
\end{multline}
\begin{multline}
\left[ \frac{\partial }{\partial t}-\frac{Z+1}{\tau }\frac{\partial }{%
\partial \mu }\left( 1-\mu ^{2}\right) \frac{\partial }{\partial \mu }%
\right] \text{ }f_{e}\left( r+\delta r,\mu ,t\right) \\ = \delta \left( \mu
-\mu _{0}-\delta \mu _{mn}\right) \delta \left( t\right) \text{,}
\end{multline}
where $\delta \left( t\right) $ and $\delta \left( \mu -\mu _{mn}\right) $
are Dirac distributions and $\delta \mu _{mn}$ the RPA kick Eq.~(\ref{step2}%
) induced by an $\left( m,n\right) $ drive. We can neglect the gradient of
the collision time as the elementary step $\delta r$ is far smaller than $a$%
, and define the electron/hole excitation: $g\left( r,\mu ,\mu
_{0},t\right) =f_{e}\left( r,\mu ,t\right) +f_{h}\left( r,\mu ,t\right) $.
This electron-hole distribution function is solution to the kinetic equation: 
\begin{multline}
\left[ \frac{\partial }{\partial t}-\frac{Z+1}{\tau }\frac{\partial }{%
\partial \mu }\left( 1-\mu ^{2}\right) \frac{\partial }{\partial \mu }%
\right] \text{ }g\left( r,\mu ,\mu _{0},t\right) \\ =\delta \mu _{mn}\frac{%
d\delta \left( \mu -\mu _{0}\right) }{d\mu _{0}}\delta \left( t\right) \text{%
.}  \label{landau}
\end{multline}
To solve this kinetic equation we expand the electron-hole excitation $g$
over the Legendre polynomials $P_{l}\left( \mu \right) $ which are the
classical basis to study electron anisotropy in plasma kinetic problems such
as the Spitzer conductivity problem or the Fisch efficiency problem. The
Dirac pitch-angle source $\delta \left( \mu -\mu _{o}\right) $ can be
expanded as 
\begin{equation}
\delta \left( \mu -\mu _{o}\right) =\sum_{l=0}^{l=+\infty }\frac{2l+1}{2}%
P_{l}\left( \mu \right) P_{l}\left( \mu _{o}\right) \text{.}  \label{comp}
\end{equation}
The Legendre polynomials are the eigenfunctions of the pitch-angle
isotropization kinetic operator: 
\begin{equation}
\left[ \frac{\partial }{\partial \mu }\left( 1-\mu ^{2}\right) \frac{%
\partial }{\partial \mu }\right] P_{l}\left( \mu \right) +l\left( l+1\right)
P_{l}\left( \mu \right) =0\text{.}  \label{eig}
\end{equation}
These two relations, Eq.~(\ref{comp},\ref{eig}), allow to solve analytically
the Landau kinetic equation Eq.~(\ref{landau}). The various anisotropic
components decay exponentially and 
\begin{widetext}
\begin{equation}
g\left( r,\mu ,\mu _{0},t\right) =H\left( t\right) \delta \mu
_{mn}\sum_{l=1}^{l=+\infty }\frac{2l+1}{2}\exp \left[ -l\left( l+1\right)
\left( Z+1\right) \frac{t}{\tau }\right] P_{l}\left( \mu \right)
P_{l}^{\prime }\left( \mu _{o}\right),  \label{gg}
\end{equation}
\end{widetext}
where $H\left( t\right) $ is the Heaviside step function such that $H\left(
t\geq 0\right) =1$ and $H\left( t<0\right) =0$ and the prime indicates a
derivative with respect to $\mu _{0}$.

In a centrally fuelled steady-state tokamak, with a single $\left( m,n\right) 
$ turbulent mode, during a time $dt$, an average number of electrons $%
dN=W_{f}dt/Q_{DT}$ jump from $r$ to $r+\delta r$. As all the electrons are
equally involved, the distribution of $\mu _{0}$ variable is flat
between $-1$ and $+1$, so the steady-state non-equilibrium pitch-angle
distribution $G_{nm}\left( r,\mu \right) $ at radius $r$ is given by the
average:
\begin{align}
G_{nm}\left( r,\mu \right) = & \int gd\mu _{0}dN \nonumber\\ = & \frac{W_{f}}{Q_{DT}}%
\int_{-1}^{+1}d\mu _{0}\int_{-\infty }^{+\infty }g\left( r,\mu ,\mu
_{0},t\right) dt.
\end{align}
The Legendre polynomials expansion Eq.~(\ref{gg}) provides the final result
as a sum of odd and even components: 
\begin{widetext}
\begin{equation}
G_{nm}\left( r,\mu \right) =\frac{W_{f}}{Q_{DT}}\frac{r\delta r}{Rv}\tau
\left[ \sum_{l=1,3,5...}\frac{n\left( 2l+1\right) }{ml\left( l+1\right)
\left( Z+1\right) }P_{l}\left( \mu \right) -\sum_{l=2,4,6...}\frac{\omega
_{mn}R\left( 2l+1\right) }{ml\left( l+1\right) \left( Z+1\right) v}%
P_{l}\left( \mu \right) \right] \text{.}
\end{equation}
\end{widetext}
The current $\delta I_{nm}\left( r\right) $ associated with steady-state
electron/hole excitations by a single $\left( n,m\right) $ mode is given by
the $v_{\Vert }=v\mu $ moment of the $G_{nm}\left( r,\mu \right) $ non-equilibrium distribution function 
\begin{align}
\delta I_{nm}\left( r,\delta r\right) = & \frac{e}{2\pi R}\int_{-1}^{+1}v\mu
G_{mn}\left( r,\mu \right) d\mu \nonumber\\ = &\frac{e}{2\pi R}\frac{\Omega W_{f}}{Q_{DT}R}%
\frac{n}{m}\frac{r\tau }{\left( Z+1\right) }\delta r\text{.}
\end{align}
If the tokamak is fuelled from the edge 
\begin{equation}
\sum_{\text{random walk}}\delta r=0\quad\textrm{and}\quad\sum_{\text{random walk}}\delta I_{mn}=0,
\end{equation}
whereas if the tokamak is
centrally fuelled
\begin{equation}
\sum_{\text{random walk}}\delta r=a\quad\textrm{and}\quad\sum_{\text{random walk}%
}\delta I_{mn}\neq 0.
\end{equation}

The expression of the full transport driven current associated with the mode 
$\left( m,n\right) $ must be averaged over a Maxwellian distribution of the
velocity $v$ with temperature $T_{e}\left( r\right) $ and then integrated
from the center to the edge ($\sum \delta r=\int dr$) with respect to the
random radial walk: 
\begin{align}
I_{nm} = &\sum_{\delta r}\left\langle \delta I_{mn}\left( r,\delta r\right)
\right\rangle _{T_{e}\left( r\right) }\nonumber \\ = &\frac{e}{2\pi \left( Z+1\right) }%
\frac{\Omega W_{f}}{Q_{DT}R^{2}}\frac{n}{m}\int_{0}^{a}drr\left\langle \tau
\left( v,r\right) \right\rangle _{T_{e}\left( r\right) }\text{.}
\end{align}
We introduce the mean collision time $\left\langle \tau \right\rangle $ as
an average over velocity and radial position according to the relation: 
\begin{align}
\left\langle \tau \right\rangle = & \frac{\int_{0}^{a}r\left\langle \tau \left(
v,r\right) \right\rangle _{T_{e}\left( r\right) }dr}{a^{2}/2}\nonumber\\ = & \frac{%
64{\varepsilon _{0}}^{2}\sqrt{2\pi m_{e}}{k_{B}}^{3/2}}{e^{4}a^{2}\log \Lambda }%
\int_{0}^{a}\frac{T_{e}\left( r\right) ^{\frac{3}{2}}}{n_{e}\left( r\right) }%
rdr\text{,}  \label{meant}
\end{align}
so that the $\left( m,n\right) $ driven current is 
\begin{equation}
\frac{I_{mn}}{W_{f}}=\frac{\left\langle \tau \right\rangle \Omega }{2\pi
\left( Z+1\right) }\frac{e}{Q_{DT}}\frac{a^{2}}{R^{2}}\frac{n}{m}\text{.}
\end{equation}
This relation assumes that a single $\left( n,m\right) $ mode is at work to
provide the radial collisionless transport of the electrons from $r=0$ to $%
r=a$. In fact, the turbulent activity of a discharge is associated with a
spectrum of $\ m$ and $n$ and we have to define a mean spectral
characteristic of the discharge to express the transport driven current.

\section{Discussion and conclusion}
\label{Sec:Sec4}

As we work within the framework of the RPA, we can sum the effects of each $%
\left( m,n\right) $ mode and neglect the interferences between the various
modes~\cite{Wesson2011,Rax2011,Kaufman1972,White2014}. As identified and discussed at the end of Sec.~\ref{Sec:Sec2}, each $\left( m,n\right) $ mode contributes to the full
quasilinear radial diffusion coefficients $\left\langle \delta
{D_{kmn}}^{2}\right\rangle $/$2\delta t$ and $\left\langle \delta
D_{kmn}\delta P_{kmn}\right\rangle $/$2\delta t$.  Specifically, Eqs.~(\ref{ql1}, \ref{ql2}, \ref{ql3}) show that the contribution of mode $(m,n)$ is proportional to ${\phi _{mn}}^{2}$. Thus we introduce a
coefficient proportional to ${\phi _{mn}}^{2}$ measuring the relative
contribution of each $\left( m,n\right) $ mode to quasilinear diffusion in $%
\left( J,D,P\right) $ space, that is to say to current generation. The final
formulae for the full transport driven current $I$ is thus given by 
\begin{equation}
\frac{I}{W_{f}}=\frac{\Omega \left\langle \tau \right\rangle }{2\pi \left(
Z+1\right) }\frac{e}{Q_{DT}}\frac{a^{2}}{R^{2}}\left\langle \frac{n}{m}%
\right\rangle \text{,}  \label{current}
\end{equation}
where we have defined the mean ratio of toroidal to poloidal mode number $%
\left\langle n/m\right\rangle $ as 
\begin{widetext}
\begin{equation}
\left\langle \frac{n}{m}\right\rangle =\sum_{m=-\infty }^{m=+\infty
}\sum_{n=-\infty }^{n=+\infty }\frac{n}{m}\int_{0}^{+\infty }kdk\frac{{\phi
_{mn}}^{2}\left( k\right) }{\sum_{m,n=-\infty }^{m,n=+\infty
}\int_{0}^{+\infty }udu{\phi _{mn}}^{2}\left( u\right) }.  \label{mnmn}
\end{equation}
\end{widetext}
There is no poloidally isotropic mode $\phi _{0n}=0$ in the spectrum and the
energy ${\phi _{mn}}^{2}$ content of each mode reflects its contribution to
radial quasilinear transport.

Equation (\ref{current}), which quantifies the transport driven current $I$, was
derived under two hypotheses. First, the interaction between an electron and electrostatic modes has been assumed to be governed by RPA quasilinear transport, as supported by the careful identification of the slow action and the fast phases of
the adiabatic motion as given in Sec.~\ref{Sec:Sec2}. Second, it has been assumed in Sec.~\ref{Sec:Sec3} that collisional relaxation is consistent with Landau kinetic theory. To the extent that these two frameworks are the standard descriptions for
mode-particle and particle-particle interactions in tokamak
physics~\cite{Wesson2011,Rax2011}, the final relation Eq.~(\ref{current}) is valid within the regime of applicability of these approaches. However, note that if, for example, anomalous
electron transport arises from magnetic turbulence along random magnetic field
lines~\cite{Rechester1978,Rax1992}, then this model of electrostatic turbulence and the effect
of transport driven current described by Eq.~(\ref{current}) are no longer
valid.

In order to provide a general simple scaling we consider that the radial
temperature and density profiles are characterized by an exponent $\gamma $
such that 
\begin{equation}
\frac{T_{e}\left( r\right) ^{\frac{3}{2}}}{n_{e}\left( r\right) }=\frac{%
{T_{0}}^{\frac{3}{2}}}{n_{0}}\left( 1-\frac{r^{2}}{a^{2}}\right) ^{\gamma }%
\text{,}  \label{prof}
\end{equation}
where $T_{0}$ is the electron temperature on the magnetic axis and $n_{0}$
the electron density on axis. With this profile Eq.~(\ref{prof}) the mean
relaxation time $\left\langle \tau \right\rangle $ Eq.~(\ref{meant}) becomes 
\begin{equation}
\left\langle \tau \right\rangle =\frac{32{\varepsilon _{0}}^{2}\sqrt{2\pi m_{e}%
}}{\left( \gamma +1\right) e^{5/2}\log \Lambda n_{0}}\left( \frac{k_{B}T_{0}%
}{e}\right) ^{\frac{3}{2}}.
\label{prof2}
\end{equation}
Plugging Eq.(\ref{prof2}) into Eq.(\ref
{current}), we get the scaling of the transport current as a function of the
plasma parameters for a centrally fuelled cylindrical thermonuclear
discharge: 
\begin{multline}
\frac{I}{W_{f}}\left[ \frac{\text{A}}{\text{W}}\right] \approx \frac{3}{%
\left( Z+1\right) \left( \gamma +1\right) \log \Lambda }\frac{a^{2}}{R^{2}}%
\left\langle \frac{n}{m}\right\rangle \left[ \frac{B}{1\text{ T}}\right]\\\times
\left[ \frac{k_{B}T_{0}/e}{1\text{ kV}}\right] ^{\frac{3}{2}}\left[ \frac{%
10^{13}\text{ cm}^{-3}}{n_{0}}\right] \text{.}  \label{current2}
\end{multline}
Equations (\ref{current},\ref{current2}) are the main original results of
this study.

The only unknown parameter in this relation is $\left\langle
n/m\right\rangle $ defined in Eq.~(\ref{mnmn}). In tokamaks, unstable modes
feeding the turbulence spectrum are localized near resonant drift surface
associated with closed helical field lines. This motivates us to assume here that $\left\langle n/m\right\rangle \sim 1/q$, where $q$ is the
mean safety factor of the discharge. However, the validity of this last hypothesis should be confirmed in future studies. Indeed, if the electrostatic spectrum were to be such that $\left\langle n/m\right\rangle \sim 0$, then the effect would be much weaker. It is worth noting here though that the coefficient involved in Eq.~(\ref
{current}) is not $\left\langle n\right\rangle $ or $\left\langle
m\right\rangle $ separately but $\left\langle n/m\right\rangle $.

For typical ITER parameters, and assuming $\left\langle n/m\right\rangle \sim
1/q $, Eq.~(\ref{current2}) predicts a transport driven current of few mega-amperes which confirms the favorable scaling already identified in
Sec.~\ref{Sec:Sec1}. This remains true even if accounting for trapped particles. Indeed, introducing the fraction of passing particles $P_{e}\left( r\right) =1-\sqrt{2r/R}$ on the
drift surface $r$, the kinetic analytical model can be extended by substituting the 
the radial average
\begin{equation}
\left\langle \tau \right\rangle =\frac{64{\varepsilon _{0}}^{2}\sqrt{2\pi m_{e}%
}{k_{B}}^{3/2}}{e^{4}a^{2}\log \Lambda }\int_{0}^{a}P_{e}\left( r\right) \frac{%
T_{e}\left( r\right) ^{\frac{3}{2}}}{n_{e}\left( r\right) }rdr\text{,}
\end{equation}
in lieu of Eq.~(\ref{meant}). With the general radial profile
\begin{equation}
P_{e}\left( r\right) \frac{T_{e}\left( r\right) ^{\frac{3}{2}}}{n_{e}\left(
r\right) }=\left( 1-\sqrt{\frac{2r}{R}}\right) \frac{{T_{0}}^{\frac{3}{2}}}{%
n_{0}}\left( 1-\frac{r^{2}}{a^{2}}\right) ^{\gamma },
\end{equation}
the $I/W_{f}$ expression in Eq.~(\ref{current2}) is then multiplied by the
correcting factor:
\begin{equation}
0<1-\frac{\left( 1+\gamma \right) \Gamma \left( \gamma +1\right) \Gamma
\left( 5/4\right) }{\Gamma \left( \gamma +9/4\right) }\sqrt{\frac{2a}{R}}<1%
\text{,}  \label{pass}
\end{equation}
where $\Gamma $ is the gamma function defined by Euler's integral $\Gamma
\left( u\right) =$ $\int_{0}^{+\infty }t^{u-1}\exp \left( -t\right) dt$.
As anticipated, this correction does not change the order of magnitude for $I$ and just
lower the cylindrical result by a factor one half to one third depending on $%
\gamma $.

It is to be noted that transport driven current suffers from a drawback
similar to the bootstrap current: the current on the magnetic axis cancels.
This transport driven current effect can be interpreted as a slight preferential
loss of electrons traveling\ in the direction of the toroidal current under
the hypothesis of a centrally fuelled discharge.

In summary, we have identified, described and analyzed the transport driven
current due to central fuelling in cylindrical and toroidal discharges. The
interplay between current generation and radial transport was explored with
a phenomenological model in Ref.~\cite{Rax1989} or within the framework of magnetic
turbulence in Ref.~\cite{Rax1999}. However, these studies did not take into account the
consequences of the quasilinear hypothesis Eq.~(\ref{step2}) and the central
fuelling hypothesis, and hence missed this effect. Ref.~\cite{Nunan1994}
reports the first observation of this effect but is restricted to 2+1/2
dimensional electromagnetic, particle-in-cell simulations. The original
analytical kinetic theory presented in this study is supported by these
early results in the collisionless regime. However, and although the first-principles mechanisms are similar, direct comparison of the
current is not possible because of the electron to ion
mass ratio used in these particle-in-cell studies.

The first-principles analytical kinetic model derived in this paper is
based on two standard assumptions: (\textit{i}) collisional
relaxation of anisotropy (current) is faster than anomalous radial transport and (\textit{ii}) tokamak kinetics is described by quasilinear and Landau equations. This suggests that the final cylindrical scaling Eq.~(\ref{current}) and
toroidal correction factor Eq.~(\ref{pass}) are robust results. On the other hand, what must be improved through
further studies is the prediction of the order
of magnitude $\left\langle n/m\right\rangle $ defined in Eq.~(\ref{mnmn}).

The main result of this quasilinear/collisional model is Eq.~(\ref{current}), which can can be summarized as follows. If the requirements
of central fuelling and typical turbulent spectrum $\left\langle
n/m\right\rangle \sim 1/q$ were to be satisfied in an ITER discharge, an additional, transport driven, current of up to
a few mega-amperes is predicted besides the bootstrap and non-inductive currents. This additional current would improve the global power
balance of a steady-state burning plasma. 

\section*{References}


\begin{thebibliography}{29}%
\makeatletter
\providecommand \@ifxundefined [1]{%
 \@ifx{#1\undefined}
}%
\providecommand \@ifnum [1]{%
 \ifnum #1\expandafter \@firstoftwo
 \else \expandafter \@secondoftwo
 \fi
}%
\providecommand \@ifx [1]{%
 \ifx #1\expandafter \@firstoftwo
 \else \expandafter \@secondoftwo
 \fi
}%
\providecommand \natexlab [1]{#1}%
\providecommand \enquote  [1]{``#1''}%
\providecommand \bibnamefont  [1]{#1}%
\providecommand \bibfnamefont [1]{#1}%
\providecommand \citenamefont [1]{#1}%
\providecommand \href@noop [0]{\@secondoftwo}%
\providecommand \href [0]{\begingroup \@sanitize@url \@href}%
\providecommand \@href[1]{\@@startlink{#1}\@@href}%
\providecommand \@@href[1]{\endgroup#1\@@endlink}%
\providecommand \@sanitize@url [0]{\catcode `\\12\catcode `\$12\catcode
  `\&12\catcode `\#12\catcode `\^12\catcode `\_12\catcode `\%12\relax}%
\providecommand \@@startlink[1]{}%
\providecommand \@@endlink[0]{}%
\providecommand \url  [0]{\begingroup\@sanitize@url \@url }%
\providecommand \@url [1]{\endgroup\@href {#1}{\urlprefix }}%
\providecommand \urlprefix  [0]{URL }%
\providecommand \Eprint [0]{\href }%
\providecommand \doibase [0]{http://dx.doi.org/}%
\providecommand \selectlanguage [0]{\@gobble}%
\providecommand \bibinfo  [0]{\@secondoftwo}%
\providecommand \bibfield  [0]{\@secondoftwo}%
\providecommand \translation [1]{[#1]}%
\providecommand \BibitemOpen [0]{}%
\providecommand \bibitemStop [0]{}%
\providecommand \bibitemNoStop [0]{.\EOS\space}%
\providecommand \EOS [0]{\spacefactor3000\relax}%
\providecommand \BibitemShut  [1]{\csname bibitem#1\endcsname}%
\let\auto@bib@innerbib\@empty
\bibitem [{\citenamefont {Wesson}\ and\ \citenamefont
  {Campbell}(2011)}]{Wesson2011}%
  \BibitemOpen
  \bibfield  {author} {\bibinfo {author} {\bibfnamefont {J.}~\bibnamefont
  {Wesson}}\ and\ \bibinfo {author} {\bibfnamefont {D.~J.}\ \bibnamefont
  {Campbell}},\ }\href@noop {} {\emph {\bibinfo {title} {Tokamaks}}}\ (\bibinfo
   {publisher} {Oxford University Press, Oxford},\ \bibinfo {year}
  {2011})\BibitemShut {NoStop}%
\bibitem [{\citenamefont {Rax}(2011)}]{Rax2011}%
  \BibitemOpen
  \bibfield  {author} {\bibinfo {author} {\bibfnamefont {J.-M.}\ \bibnamefont
  {Rax}},\ }\href@noop {} {\emph {\bibinfo {title} {Physique des tokamaks}}}\
  (\bibinfo  {publisher} {Éd. de l'École polytechnique},\ \bibinfo
  {address} {Paris},\ \bibinfo {year} {2011})\BibitemShut {NoStop}%
\bibitem [{\citenamefont {Fisch}(2014)}]{Fisch2014}%
  \BibitemOpen
  \bibfield  {author} {\bibinfo {author} {\bibfnamefont {N.~J.}\ \bibnamefont
  {Fisch}},\ }\href {\doibase 10.13182/fst13-670} {\bibfield  {journal}
  {\bibinfo  {journal} {Fusion Sci. Technol.}\ }\textbf {\bibinfo {volume}
  {65}},\ \bibinfo {pages} {1} (\bibinfo {year} {2014})}\BibitemShut {NoStop}%
\bibitem [{\citenamefont {Fisch}(1978)}]{Fisch1978}%
  \BibitemOpen
  \bibfield  {author} {\bibinfo {author} {\bibfnamefont {N.~J.}\ \bibnamefont
  {Fisch}},\ }\href {\doibase 10.1103/physrevlett.41.873} {\bibfield  {journal}
  {\bibinfo  {journal} {Phys. Rev. Lett.}\ }\textbf {\bibinfo {volume} {41}},\
  \bibinfo {pages} {873} (\bibinfo {year} {1978})}\BibitemShut {NoStop}%
\bibitem [{\citenamefont {Fisch}(1987)}]{Fisch1987}%
  \BibitemOpen
  \bibfield  {author} {\bibinfo {author} {\bibfnamefont {N.~J.}\ \bibnamefont
  {Fisch}},\ }\href {\doibase 10.1103/revmodphys.59.175} {\bibfield  {journal}
  {\bibinfo  {journal} {Rev. Mod. Phys.}\ }\textbf {\bibinfo {volume} {59}},\
  \bibinfo {pages} {175} (\bibinfo {year} {1987})}\BibitemShut {NoStop}%
\bibitem [{\citenamefont {Bickerton}, \citenamefont {Connor},\ and\
  \citenamefont {Taylor}(1971)}]{Bickerton1971}%
  \BibitemOpen
  \bibfield  {author} {\bibinfo {author} {\bibfnamefont {R.~J.}\ \bibnamefont
  {Bickerton}}, \bibinfo {author} {\bibfnamefont {J.~W.}\ \bibnamefont
  {Connor}}, \ and\ \bibinfo {author} {\bibfnamefont {J.~B.}\ \bibnamefont
  {Taylor}},\ }\href {\doibase 10.1038/physci229110a0} {\bibfield  {journal}
  {\bibinfo  {journal} {Nature Phys. Sci.}\ }\textbf {\bibinfo {volume}
  {229}},\ \bibinfo {pages} {110} (\bibinfo {year} {1971})}\BibitemShut
  {NoStop}%
\bibitem [{\citenamefont {Kadomtsev}\ and\ \citenamefont
  {Shafranov}(1971)}]{Kadomtsev1971}%
  \BibitemOpen
  \bibfield  {author} {\bibinfo {author} {\bibfnamefont {B.}~\bibnamefont
  {Kadomtsev}}\ and\ \bibinfo {author} {\bibfnamefont {V.}~\bibnamefont
  {Shafranov}},\ }in\ \href@noop {} {\emph {\bibinfo {booktitle} {Proceedings
  of the Fourth International Conference on Plasma Physics and Controlled
  Nuclear Fusion Research, Madison, WI, USA}}},\ Vol.~\bibinfo {volume} {2}\
  (\bibinfo {organization} {International Atomic Energy Agency, Vienna},\
  \bibinfo {year} {1971})\ p.\ \bibinfo {pages} {479}\BibitemShut {NoStop}%
\bibitem [{\citenamefont {Nunan}\ and\ \citenamefont
  {Dawson}(1994)}]{Nunan1994}%
  \BibitemOpen
  \bibfield  {author} {\bibinfo {author} {\bibfnamefont {W.~J.}\ \bibnamefont
  {Nunan}}\ and\ \bibinfo {author} {\bibfnamefont {J.~M.}\ \bibnamefont
  {Dawson}},\ }\href {\doibase 10.1103/physrevlett.73.1628} {\bibfield
  {journal} {\bibinfo  {journal} {Phys. Rev. Lett.}\ }\textbf {\bibinfo
  {volume} {73}},\ \bibinfo {pages} {1628} (\bibinfo {year}
  {1994})}\BibitemShut {NoStop}%
\bibitem [{\citenamefont {Ma}\ and\ \citenamefont {Dawson}(1994)}]{Ma1994}%
  \BibitemOpen
  \bibfield  {author} {\bibinfo {author} {\bibfnamefont {S.}~\bibnamefont
  {Ma}}\ and\ \bibinfo {author} {\bibfnamefont {J.~M.}\ \bibnamefont
  {Dawson}},\ }\href {\doibase 10.1063/1.870593} {\bibfield  {journal}
  {\bibinfo  {journal} {Phys. Plasmas}\ }\textbf {\bibinfo {volume} {1}},\
  \bibinfo {pages} {2661} (\bibinfo {year} {1994})}\BibitemShut {NoStop}%
\bibitem [{\citenamefont {Galeev}\ and\ \citenamefont
  {Sagdeev}(1968)}]{Galeev1968}%
  \BibitemOpen
  \bibfield  {author} {\bibinfo {author} {\bibfnamefont {A.~A.}\ \bibnamefont
  {Galeev}}\ and\ \bibinfo {author} {\bibfnamefont {R.~Z.}\ \bibnamefont
  {Sagdeev}},\ }\href@noop {} {\bibfield  {journal} {\bibinfo  {journal} {Sov.
  Phys. JETP}\ }\textbf {\bibinfo {volume} {26}},\ \bibinfo {pages} {233}
  (\bibinfo {year} {1968})}\BibitemShut {NoStop}%
\bibitem [{\citenamefont {Kaufman}(1972)}]{Kaufman1972}%
  \BibitemOpen
  \bibfield  {author} {\bibinfo {author} {\bibfnamefont {A.~N.}\ \bibnamefont
  {Kaufman}},\ }\href {\doibase 10.1063/1.1694031} {\bibfield  {journal}
  {\bibinfo  {journal} {Phys. Fluids}\ }\textbf {\bibinfo {volume} {15}},\
  \bibinfo {pages} {1063} (\bibinfo {year} {1972})}\BibitemShut {NoStop}%
\bibitem [{\citenamefont {Robiche}\ and\ \citenamefont
  {Rax}(2004)}]{Robiche2004}%
  \BibitemOpen
  \bibfield  {author} {\bibinfo {author} {\bibfnamefont {J.}~\bibnamefont
  {Robiche}}\ and\ \bibinfo {author} {\bibfnamefont {J.~M.}\ \bibnamefont
  {Rax}},\ }\href {\doibase 10.1103/physreve.70.046405} {\bibfield  {journal}
  {\bibinfo  {journal} {Phys. Rev. E}\ }\textbf {\bibinfo {volume} {70}},\
  \bibinfo {pages} {046405} (\bibinfo {year} {2004})}\BibitemShut {NoStop}%
\bibitem [{\citenamefont {Ware}(1970)}]{Ware1970}%
  \BibitemOpen
  \bibfield  {author} {\bibinfo {author} {\bibfnamefont {A.~A.}\ \bibnamefont
  {Ware}},\ }\href {\doibase 10.1103/physrevlett.25.916} {\bibfield  {journal}
  {\bibinfo  {journal} {Phys. Rev. Lett.}\ }\textbf {\bibinfo {volume} {25}},\
  \bibinfo {pages} {916} (\bibinfo {year} {1970})}\BibitemShut {NoStop}%
\bibitem [{\citenamefont {Rax}\ and\ \citenamefont {Moreau}(1989)}]{Rax1989}%
  \BibitemOpen
  \bibfield  {author} {\bibinfo {author} {\bibfnamefont {J.~M.}\ \bibnamefont
  {Rax}}\ and\ \bibinfo {author} {\bibfnamefont {D.}~\bibnamefont {Moreau}},\
  }\href {\doibase 10.1088/0029-5515/29/10/009} {\bibfield  {journal} {\bibinfo
   {journal} {Nucl. Fusion}\ }\textbf {\bibinfo {volume} {29}},\ \bibinfo
  {pages} {1751} (\bibinfo {year} {1989})}\BibitemShut {NoStop}%
\bibitem [{\citenamefont {Rax}(2014)}]{Rax2014}%
  \BibitemOpen
  \bibfield  {author} {\bibinfo {author} {\bibfnamefont {J.~M.}\ \bibnamefont
  {Rax}},\ }\href {\doibase 10.13182/fst13-634} {\bibfield  {journal} {\bibinfo
   {journal} {Fusion Sci. Technol.}\ }\textbf {\bibinfo {volume} {65}},\
  \bibinfo {pages} {10} (\bibinfo {year} {2014})}\BibitemShut {NoStop}%
\bibitem [{\citenamefont {Fisch}\ and\ \citenamefont {Rax}(1992)}]{Fisch1992}%
  \BibitemOpen
  \bibfield  {author} {\bibinfo {author} {\bibfnamefont {N.~J.}\ \bibnamefont
  {Fisch}}\ and\ \bibinfo {author} {\bibfnamefont {J.-M.}\ \bibnamefont
  {Rax}},\ }\href {\doibase 10.1103/PhysRevLett.69.612} {\bibfield  {journal}
  {\bibinfo  {journal} {Phys. Rev. Lett.}\ }\textbf {\bibinfo {volume} {69}},\
  \bibinfo {pages} {612} (\bibinfo {year} {1992})}\BibitemShut {NoStop}%
\bibitem [{\citenamefont {Fisch}\ and\ \citenamefont {Rax}(1993)}]{Fisch1993}%
  \BibitemOpen
  \bibfield  {author} {\bibinfo {author} {\bibfnamefont {N.~J.}\ \bibnamefont
  {Fisch}}\ and\ \bibinfo {author} {\bibfnamefont {J.-M.}\ \bibnamefont
  {Rax}},\ }\href {\doibase 10.1063/1.860809} {\bibfield  {journal} {\bibinfo
  {journal} {Phys. Fluids B}\ }\textbf {\bibinfo {volume} {5}},\ \bibinfo
  {pages} {1754} (\bibinfo {year} {1993})}\BibitemShut {NoStop}%
\bibitem [{\citenamefont {Fisch}\ and\ \citenamefont
  {Herrmann}(1994)}]{Fisch1994}%
  \BibitemOpen
  \bibfield  {author} {\bibinfo {author} {\bibfnamefont {N.}~\bibnamefont
  {Fisch}}\ and\ \bibinfo {author} {\bibfnamefont {M.}~\bibnamefont
  {Herrmann}},\ }\href {\doibase 10.1088/0029-5515/34/12/i01} {\bibfield
  {journal} {\bibinfo  {journal} {Nucl. Fusion}\ }\textbf {\bibinfo {volume}
  {34}},\ \bibinfo {pages} {1541} (\bibinfo {year} {1994})}\BibitemShut
  {NoStop}%
\bibitem [{\citenamefont {Heikkinen}\ and\ \citenamefont
  {Sipilä}(1995)}]{Heikkinen1995}%
  \BibitemOpen
  \bibfield  {author} {\bibinfo {author} {\bibfnamefont {J.~A.}\ \bibnamefont
  {Heikkinen}}\ and\ \bibinfo {author} {\bibfnamefont {S.~K.}\ \bibnamefont
  {Sipilä}},\ }\href {\doibase 10.1063/1.871072} {\bibfield  {journal}
  {\bibinfo  {journal} {Phys. Plasmas}\ }\textbf {\bibinfo {volume} {2}},\
  \bibinfo {pages} {3724} (\bibinfo {year} {1995})}\BibitemShut {NoStop}%
\bibitem [{\citenamefont {Herrmann}\ and\ \citenamefont
  {Fisch}(1997)}]{Herrmann1997}%
  \BibitemOpen
  \bibfield  {author} {\bibinfo {author} {\bibfnamefont {M.~C.}\ \bibnamefont
  {Herrmann}}\ and\ \bibinfo {author} {\bibfnamefont {N.~J.}\ \bibnamefont
  {Fisch}},\ }\href {\doibase 10.1103/physrevlett.79.1495} {\bibfield
  {journal} {\bibinfo  {journal} {Phys. Rev. Lett.}\ }\textbf {\bibinfo
  {volume} {79}},\ \bibinfo {pages} {1495} (\bibinfo {year}
  {1997})}\BibitemShut {NoStop}%
\bibitem [{\citenamefont {Ochs}, \citenamefont {Bertelli},\ and\ \citenamefont
  {Fisch}(2015)}]{Ochs2015}%
  \BibitemOpen
  \bibfield  {author} {\bibinfo {author} {\bibfnamefont {I.~E.}\ \bibnamefont
  {Ochs}}, \bibinfo {author} {\bibfnamefont {N.}~\bibnamefont {Bertelli}}, \
  and\ \bibinfo {author} {\bibfnamefont {N.~J.}\ \bibnamefont {Fisch}},\ }\href
  {\doibase 10.1063/1.4935123} {\bibfield  {journal} {\bibinfo  {journal}
  {Phys. Plasmas}\ }\textbf {\bibinfo {volume} {22}},\ \bibinfo {pages}
  {112103} (\bibinfo {year} {2015})}\BibitemShut {NoStop}%
\bibitem [{\citenamefont {Cook}, \citenamefont {Dendy},\ and\ \citenamefont
  {Chapman}(2017)}]{Cook2017}%
  \BibitemOpen
  \bibfield  {author} {\bibinfo {author} {\bibfnamefont {J.}~\bibnamefont
  {Cook}}, \bibinfo {author} {\bibfnamefont {R.}~\bibnamefont {Dendy}}, \ and\
  \bibinfo {author} {\bibfnamefont {S.}~\bibnamefont {Chapman}},\ }\href
  {\doibase 10.1103/physrevlett.118.185001} {\bibfield  {journal} {\bibinfo
  {journal} {Phys. Rev. Lett.}\ }\textbf {\bibinfo {volume} {118}},\ \bibinfo
  {pages} {185001} (\bibinfo {year} {2017})}\BibitemShut {NoStop}%
\bibitem [{\citenamefont {Rax}, \citenamefont {Gueroult},\ and\ \citenamefont
  {Fisch}(2017)}]{Rax2017}%
  \BibitemOpen
  \bibfield  {author} {\bibinfo {author} {\bibfnamefont {J.~M.}\ \bibnamefont
  {Rax}}, \bibinfo {author} {\bibfnamefont {R.}~\bibnamefont {Gueroult}}, \
  and\ \bibinfo {author} {\bibfnamefont {N.~J.}\ \bibnamefont {Fisch}},\
  }\href {\doibase 10.1063/1.4977919} {\bibfield  {journal} {\bibinfo
  {journal} {Phys. Plasmas}\ }\textbf {\bibinfo {volume} {24}},\ \bibinfo
  {pages} {032504} (\bibinfo {year} {2017})}\BibitemShut {NoStop}%
\bibitem [{\citenamefont {White}(2014)}]{White2014}%
  \BibitemOpen
  \bibfield  {author} {\bibinfo {author} {\bibfnamefont {R.~B.}\ \bibnamefont
  {White}},\ }\href@noop {} {\emph {\bibinfo {title} {The Theory of Toroidally
  Confined Plasmas}}},\ \bibinfo {edition} {3rd}\ ed.\ (\bibinfo  {publisher}
  {Imperial College Press},\ \bibinfo {year} {2014})\BibitemShut {NoStop}%
\bibitem [{\citenamefont {Watson}(1980)}]{Watson1980}%
  \BibitemOpen
  \bibfield  {author} {\bibinfo {author} {\bibfnamefont {G.~N.}\ \bibnamefont
  {Watson}},\ }\href@noop {} {\emph {\bibinfo {title} {A Treatise on the Theory
  of Bessel Functions}}}\ (\bibinfo  {publisher} {Cambridge University Press,
  New York, NY},\ \bibinfo {year} {1980})\BibitemShut {NoStop}%
\bibitem [{\citenamefont {Rax}(1988)}]{Rax1988}%
  \BibitemOpen
  \bibfield  {author} {\bibinfo {author} {\bibfnamefont {J.~M.}\ \bibnamefont
  {Rax}},\ }\href {\doibase 10.1063/1.866739} {\bibfield  {journal} {\bibinfo
  {journal} {Phys. Fluids}\ }\textbf {\bibinfo {volume} {31}},\ \bibinfo
  {pages} {1111} (\bibinfo {year} {1988})}\BibitemShut {NoStop}%
\bibitem [{\citenamefont {Rechester}\ and\ \citenamefont
  {Rosenbluth}(1978)}]{Rechester1978}%
  \BibitemOpen
  \bibfield  {author} {\bibinfo {author} {\bibfnamefont {A.~B.}\ \bibnamefont
  {Rechester}}\ and\ \bibinfo {author} {\bibfnamefont {M.~N.}\ \bibnamefont
  {Rosenbluth}},\ }\href {\doibase 10.1103/physrevlett.40.38} {\bibfield
  {journal} {\bibinfo  {journal} {Phys. Rev. Lett.}\ }\textbf {\bibinfo
  {volume} {40}},\ \bibinfo {pages} {38} (\bibinfo {year} {1978})}\BibitemShut
  {NoStop}%
\bibitem [{\citenamefont {Rax}\ and\ \citenamefont {White}(1992)}]{Rax1992}%
  \BibitemOpen
  \bibfield  {author} {\bibinfo {author} {\bibfnamefont {J.~M.}\ \bibnamefont
  {Rax}}\ and\ \bibinfo {author} {\bibfnamefont {R.~B.}\ \bibnamefont
  {White}},\ }\href {\doibase 10.1103/physrevlett.68.1523} {\bibfield
  {journal} {\bibinfo  {journal} {Phys. Rev. Lett.}\ }\textbf {\bibinfo
  {volume} {68}},\ \bibinfo {pages} {1523} (\bibinfo {year}
  {1992})}\BibitemShut {NoStop}%
\bibitem [{\citenamefont {Rax}, \citenamefont {Robiche},\ and\ \citenamefont
  {Kostyukov}(1999)}]{Rax1999}%
  \BibitemOpen
  \bibfield  {author} {\bibinfo {author} {\bibfnamefont {J.~M.}\ \bibnamefont
  {Rax}}, \bibinfo {author} {\bibfnamefont {J.}~\bibnamefont {Robiche}}, \ and\
  \bibinfo {author} {\bibfnamefont {I.}~\bibnamefont {Kostyukov}},\ }\href
  {\doibase 10.1063/1.873563} {\bibfield  {journal} {\bibinfo  {journal} {Phys.
  Plasmas}\ }\textbf {\bibinfo {volume} {6}},\ \bibinfo {pages} {3233}
  (\bibinfo {year} {1999})}\BibitemShut {NoStop}%
\end{thebibliography}

%

\end{document}